\documentclass[journal]{IEEEtran}

\ifCLASSINFOpdf
\else
   \usepackage[dvips]{graphicx}
\fi
\usepackage{url}
\usepackage{amsmath,amssymb,amsfonts, multirow}
\usepackage{graphicx}

\usepackage{bm}
\usepackage{color}

\begin{document}

\title{Frame-level Temporal Difference Learning for Partial Deepfake Speech Detection
\thanks{$^*$X.-P. Zhang is the corresponding author.}}

\author{Menglu Li, \IEEEmembership{Student Member, IEEE}, Xiao-Ping Zhang$^{*}$, \IEEEmembership{Fellow, IEEE}, and Lian Zhao, \IEEEmembership{Fellow, IEEE}
% \thanks{This paragraph of the first footnote will contain the date on which you submitted your paper for review. It will also contain support information, including sponsor and financial support acknowledgment. For example, ``This work was supported in part by the U.S. Department of Commerce under Grant BS123456.'' }
% \thanks{The next few paragraphs should contain the authors' current affiliations, including current address and e-mail. For example, F. A. Author is with the National Institute of Standards and Technology, Boulder, CO 80305 USA (e-mail: author@boulder.nist.gov).}
\thanks{Menglu Li is with the Department of Electrical, Computer and Biomedical Engineering, Toronto Metropolitan University, Toronto, ON, Canada (e-mail:menglu.li@torontomu.ca).} 
\thanks{X.-P. Zhang is with Shenzhen Key Laboratory of Ubiquitous Data Enabling, Tsinghua Shenzhen International Graduate School, Tsinghua University, and with the Department of Electrical, Computer and Biomedical Engineering, Toronto Metropolitan University, Toronto, ON, Canada. (e-mail: xpzhang@ieee.org).}
\thanks{Lian Zhao is with the Department of Electrical, Computer and Biomedical Engineering, Toronto Metropolitan University, Toronto, ON, Canada (e-mail:l5zhao@torontomu.ca).}}

\markboth{Journal of \LaTeX\ Class Files, Vol. 14, No. 8, August 2015}
{Shell \MakeLowercase{\textit{et al.}}: Bare Demo of IEEEtran.cls for IEEE Journals}
\maketitle

\begin{abstract}
Detecting partial deepfake speech is essential due to its potential for subtle misinformation. However, existing methods depend on costly frame-level annotations during training,  limiting real-world scalability. Also, they focus on detecting transition artifacts between bonafide and deepfake segments. As deepfake generation techniques increasingly smooth these transitions, detection has become more challenging. To address this, our work introduces a new perspective by analyzing frame-level temporal differences and reveals that deepfake speech exhibits erratic directional changes and unnatural local transitions compared to bonafide speech.  Based on this finding, we propose a Temporal Difference Attention Module (TDAM) that redefines partial deepfake detection as identifying unnatural temporal variations, without relying on explicit boundary annotations. A dual-level hierarchical difference representation captures temporal irregularities at both fine and coarse scales, while adaptive average pooling preserves essential patterns across variable-length inputs to minimize information loss. Our TDAM-AvgPool model achieves state-of-the-art performance, with an EER of 0.59\% on the PartialSpoof dataset and 0.03\% on the HAD dataset, which significantly outperforms the existing methods without requiring frame-level supervision.
\end{abstract}

\begin{IEEEkeywords}
Deepfake speech detection,  partial speech deepfake, anti-spoofing.
\end{IEEEkeywords}

\IEEEpeerreviewmaketitle

\section{Introduction}
The rise of speech deepfakes presents significant societal risks, driving efforts like the ASVspoof challenges \cite{liu2023asvspoof} to develop robust detection systems \cite{li24oa_interspeech, chen2023graph, tak21_interspeech, li2023robust}. However, these challenges have not addressed partial deepfakes, where deepfake segments are embedded within bonafide speech \cite{zhang2022partialspoof}.

Detecting partial deepfakes is challenging, as traditional methods often fail to localize short manipulated segments within mostly genuine utterances \cite{li2025survey}. Existing models often rely on frame-level labels, but obtaining such labels at scale can be challenging in practical scenarios. Moreover, current approaches largely focus on detecting transition artifacts introduced by concatenating bonafide and fake segments. However, recent generation techniques, such as spectral matching and waveform Overlap-Add \cite{zhang2022partialspoof, negroni24_asvspoof}, have smoothed transition artifacts, making boundary-based detection harder.

\begin{figure}[t]
  \centering
  \includegraphics[width=0.48\textwidth]{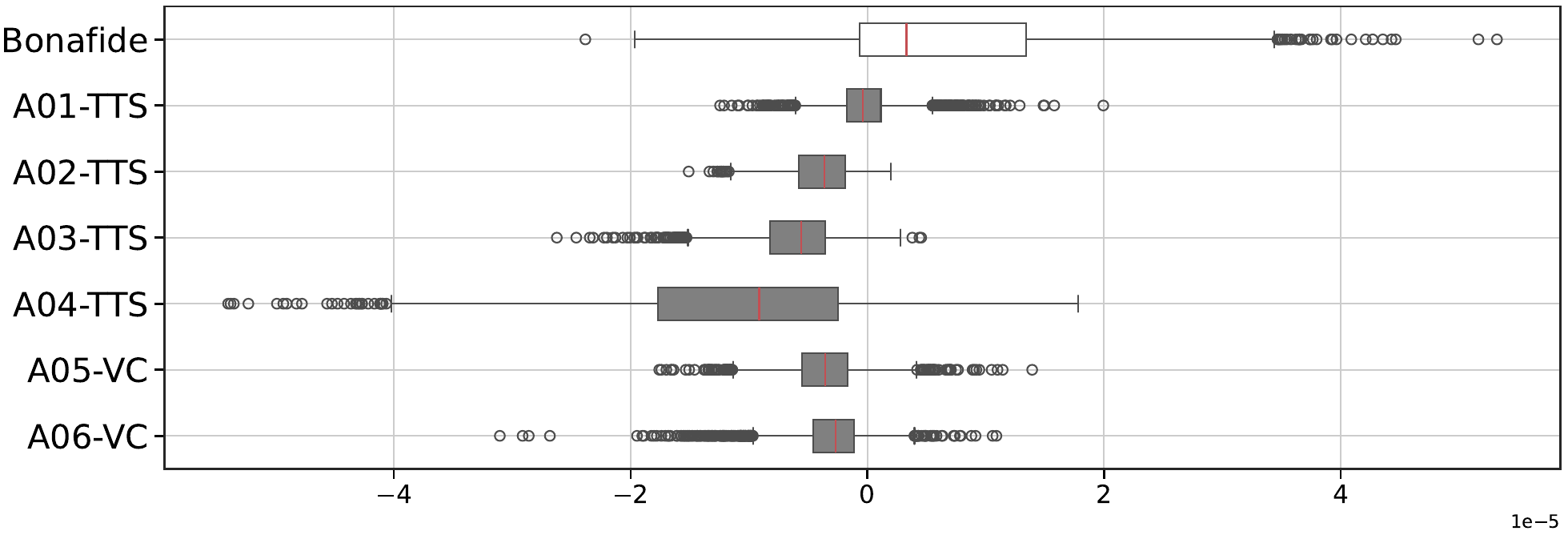}
  \caption{Comparison of mean value for raw frame-to-frame differences in wav2vec2 features across bonafide (box in white) and six different deepfake generation methods (boxes in grey) in the ASVspoof2019-LA training set.}
\label{fig:mfcc}
\end{figure} 

To address these challenges, we propose a Temporal Difference Attention Module (TDAM) that captures unnatural temporal fluctuations across utterances rather than relying on explicit boundary detection. Deepfake speech, generated by neural vocoders, often exhibits irregular spectral and directional variations between adjacent frames. As shown in Fig.~\ref{fig:mfcc}, bonafide speech exhibits a larger positive mean in raw frame-to-frame differences of wav2vec2 features, reflecting natural prosodic progression over time. In contrast, deepfake speech shows mean values near zero or negative, which suggests reduced directional consistency and more oscillatory behaviour. This indicates deepfake speech lacks natural temporal dynamics due to over-smoothing or missing long-term prosodic coherence. These findings show that directional frame-to-frame differences serve as a strong indicator for detecting deepfakes.

Building on this insight, our model leverages pre-trained embeddings to extract high-level representations and detects unnatural temporal variations through a dual-level hierarchical structure. By analyzing frame-level directional changes across the utterance, TDAM effectively identifies deepfake artifacts without relying on precise boundary detection or frame-level annotations. Trained with only utterance-level labels, it achieves state-of-the-art performance on partial deepfakes. Furthermore, our approach effectively handles varied-length inputs using average pooling, which minimizes information loss and ensures greater detection robustness.

\section{Related Work}
Early deepfake speech detection focused on fully manipulated utterances. Initial approaches used handcrafted features \cite{li2022comparative, zhang2024robust}, while recent methods leverage deep neural networks and self-supervised (SSL) models \cite{zhang2025phoneme, xie2023domain}, constructing end-to-end architectures \cite{jung2022aasist, hua2021towards} for utterance-level classification. However, these models struggle to identify partial manipulations. Partial deepfake detection methods typically fall into three categories. Frame-level approaches \cite{rahman2022detecting, martin2023vicomtech, liu2023transsionadd} analyze short segments independently but suffer from short-frame misidentification and require costly frame-level annotations. Multi-task learning \cite{li2023convolutional, li2023multi, zhang21_asvspoof} combines frame- and utterance-level tasks but increases training complexity. Boundary detection \cite{lv2022fake, cai2024integrating} targets transition regions between bonafide and fake segments, assuming detectable boundaries exist \cite{liu24m_interspeech}.
Motivated by temporal inconsistency detection in video deepfakes \cite{gu2021spatiotemporal}, we introduce a novel strategy that analyzes frame-wise feature differences to capture subtle temporal inconsistencies both within and beyond transition regions. Unlike prior methods, it achieves robust detection without relying on frame-level supervision.

\section{Temporal Difference Feature Analysis}
We analyze temporal difference features from the final layer of the wav2vec2-XLSR model to reveal distinctions in frame-level dynamics between bonafide and deepfake speech. We hypothesize that deepfakes exhibit abnormal directional patterns over time due to generation artifacts such as over-smoothing or unnatural prosody.

Given frame-wise embeddings ${\mathbf{x}_1, \mathbf{x}_2, \ldots, \mathbf{x}_T}$, we compute normalized direction vectors between consecutive frames as:
\begin{equation}
\Delta \mathbf{x}_t = \frac{\mathbf{x}_{t+1} - \mathbf{x}_t}{\|\mathbf{x}_{t+1} - \mathbf{x}_t\|}, \quad t = 1, \ldots, T-1.
\end{equation}

These vectors represent the directional flow of speech dynamics, independent of magnitude. To quantify local directional consistency, we then calculate cosine similarity between adjacent direction vectors:
\begin{equation}
\cos(\theta_t) = \frac{\Delta \mathbf{x}_t \cdot \Delta \mathbf{x}_{t+1}}{\|\Delta \mathbf{x}_t\| \|\Delta \mathbf{x}_{t+1}\|}, \quad t = 1, \ldots, T-2.
\end{equation}
For each utterance, we derive the mean and standard deviation of cosine similarity, denoted as $\mu$ and $\sigma$, respectively.

\begin{figure}
  \centering
  \includegraphics[width=0.49\textwidth]{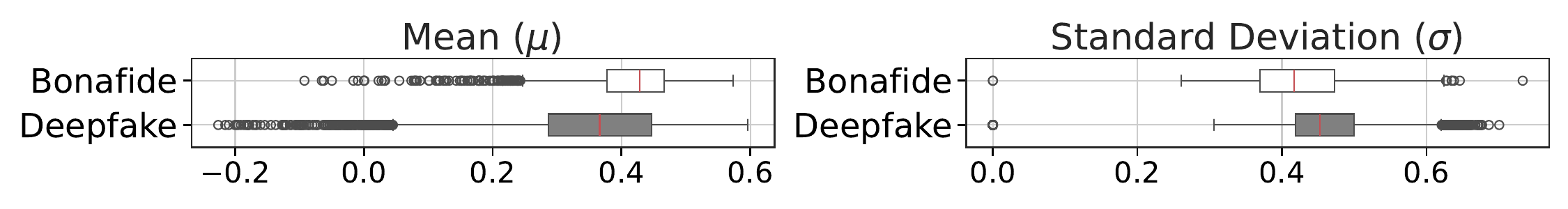}
  \caption{Comparison of mean and standard deviation of cosine similarity from wav2vec2-XLSR features for bonafide (box in white) and deepfakes (boxes in grey) on the ASVspoof2019-LA training set.}
\label{fig:cosine_stats}
\end{figure}

Applied to the ASVspoof2019-LA (LA) training set \cite{todisco19_interspeech}, the results in Fig.\ref{fig:cosine_stats} show that bonafide speech consistently yields higher $\mu$ and lower $\sigma$ than deepfakes. This indicates that bonafide speech progresses smoothly and naturally, whereas deepfake speech undergoes abrupt and erratic directional changes. The elevated $\sigma$ values in deepfakes further highlight unnatural local transitions between frames, which can be attributed to the poor modelling of prosody. These trends also align with observations in Fig.\ref{fig:mfcc}.

Our findings demonstrate that temporal difference features, particularly including directional variation information, are highly effective in distinguishing deepfake from bonafide speech. The lack of directional smoothness in deepfake utterances provides a robust cue for detection and motivates its use in model design.

\begin{figure*}[th]
  \centering
  \includegraphics[width=0.86\textwidth]{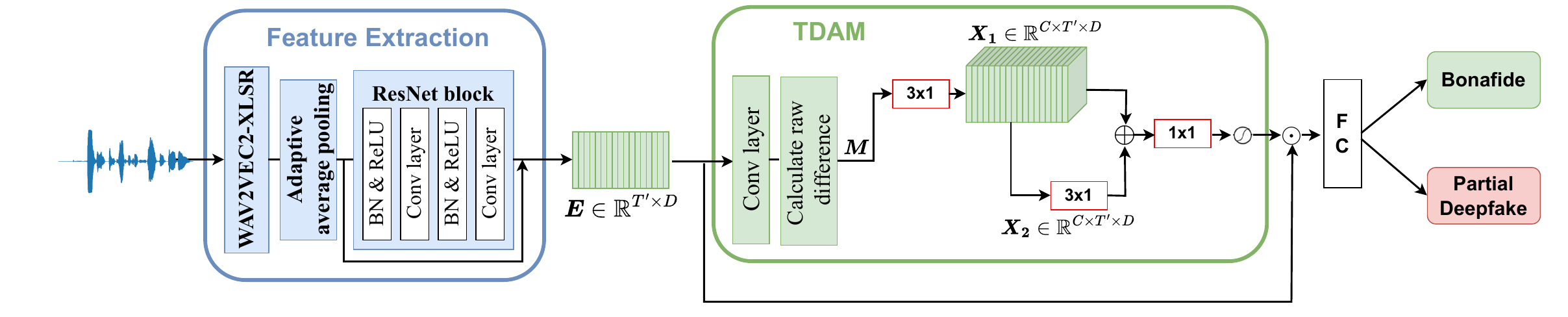}
  \caption{The architecture of the proposed detection model, TDAM-AvgPool. The model includes a feature extractor with adaptive average pooling and ResNet block, followed by the Temporal Difference Attention Module (TDAM), which is highlighted in the green box.}
\label{fig:OVERVIEW}
\end{figure*}

\section{Proposed Model}
%We formulate partial deepfake speech detection as a frame-level binary classification task. 
Figure~\ref{fig:OVERVIEW} illustrates the overall architecture of our detection model. The core component, Temporal Difference Attention Module (TDAM), captures unnatural temporal variations across adjacent frames through a dual-level design. The model also supports variable-length inputs to minimize information loss. The following sections provide details of the feature extraction process and the TDAM architecture.

\subsection{Feature Extraction for Varied-length Inputs}
Our feature extraction module converts raw waveforms into frame-level embeddings while handling variable-length inputs. Unlike traditional methods that rely on trimming or padding, which may discard critical deepfake segments, we incorporate adaptive average pooling to preserve information and enable batch processing.

We use a pre-trained wav2vec2-XLSR model \cite{babu22_interspeech} to extract high-level acoustic and temporal features directly from raw waveforms, which eliminates extensive pre-processing. All parameters are fine-tuned on labelled data to adapt to the detection task. Given an input waveform $x$, the SSL-based extractor outputs $\boldsymbol{F} \in \mathbb{R}^{T \times D}$, where $T$ varies with speech duration and $D$ is the embedding dimension. To obtain a fixed-length representation, we apply adaptive average pooling by partitioning the $T$ frames into $T'$ segments. If $T$ is not divisible by $T'$, extra frames are evenly distributed across the segments. We define $S_{t'}$ be the frame indices in the $t'$-th segment, where $t' \in \{1, 2, \dots, T'\}$. The pooled embedding $\boldsymbol{F}' \in \mathbb{R}^{T' \times D}$ is then computed as:
\begin{equation}
\boldsymbol{F}'[t'] = \frac{1}{|S_{t'}|} \sum_{t \in S_{t'}} \boldsymbol{F}[t], 
  \label{equation:eq1}
\end{equation} 
where $|S_{t'}|$ is the number of frames in $S_{t'}$. If $T < T'$, zero-padding is applied. Finally, $\boldsymbol{F}'$ is passed through two fully connected layers and a pre-activation 1D ResNet block to produce the refined embedding $\boldsymbol{E} \in \mathbb{R}^{T' \times D}$.

\subsection{Temporal Difference Attention Module}

To detect deepfake artifacts beyond transition boundaries, we propose the TDAM to capture subtle temporal irregularities by analyzing frame-to-frame feature variations with directional cues. It first constructs a difference map across adjacent frames to capture local changes, then models both fine-grained and long-range deviations using a dual-level difference representation that highlights unnatural patterns throughout the utterance.

Instead of directly using raw differences, we first apply a 1D convolution along the temporal dimension of $\boldsymbol{E}$ to aggregate contextual information and enhance robustness to noise, which produces a transformed embedding $\boldsymbol{E^{\text{conv}}} \in \mathbb{R}^{T' \times D}$. The directional difference map is then computed as:
\begin{equation}
  \boldsymbol{M}[t] = \boldsymbol{E^{\text{conv}}}[t+1] - \boldsymbol{E}[t], \quad \text{for } t = 1, \dots, T'-1.
  \label{equation:eq2}
\end{equation}  
with a zero vector of length $D$ appended to the end of $\boldsymbol{M}$ to maintain shape consistency.

TDAM hierarchically captures temporal irregularities in the embedding space through two levels of difference representations. First, $\boldsymbol{M}$ is reshaped as a time-spatial feature map with an added channel dimension. A 2D convolution expands the channel size from 1 to 32 and produces $\boldsymbol{X_1} \in \mathbb{R}^{C \times T' \times D}$. This first-level representation captures fine-grained local inconsistencies, such as micro-prosodic fluctuations, with a narrow receptive field. For a broader temporal perspective, $\boldsymbol{X_1}$ is downsampled to a channel size of 4, proceed with another convolution to expand the receptive field, and then upsampled back to 32 channels to produce $\boldsymbol{X_2} \in \mathbb{R}^{C \times T' \times D}$. This second representation models longer-term motion and directional trends, which complement local deviations learned at the first level. These two levels of representations are fused via element-wise addition, followed by a sigmoid activation to generate frame-level confidence scores. The scores weight the original embedding, $\boldsymbol{E}$, and emphasize regions where temporal progression deviates from expected natural dynamics:
\begin{equation}
  \boldsymbol{Y} = \sigma[l * (\boldsymbol{X_1} + \boldsymbol{X_2}) ] \odot \boldsymbol{E},
  \label{equation:3}
\end{equation}
where $l$ is a 2D convolution with a kernel size of 1. Through this hierarchical multi-scale design, TDAM effectively captures abrupt directional transitions and over-smoothing in deepfake speech without relying on frame-level labels or spectral features, and generates a weighted map that highlights unnatural temporal variations across the entire utterance.

A fully connected layer with softmax is applied to $\boldsymbol{Y}$ to produce frame-level predictions, which are averaged across time to obtain the utterance-level detection result.

\section{Experiment Setup}
\subsection{Dataset and Evaluation Metrics}
We evaluate our model on the PartialSpoof (PS) \cite{zhang2022partialspoof} and Half-Truth (HAD) \cite{yi21_interspeech} datasets. PS contains utterances with varying proportions of deepfake segments and includes 11 unseen attacks in its evaluation set, which is widely used for testing model generalization. HAD is a Chinese partial deepfake dataset with accent variability. We report Equal Error Rate (EER) for PS, and both EER and Area Under the Curve (AUC) for HAD, with lower EER and higher AUC indicating better performance.

\subsection{Training Configurations}
We use the pre-trained wav2vec2-XLSR model to extract 1024-dimensional embeddings every 20 ms, which are reduced to 64 dimensions via two linear layers with a dropout rate of 0.2. Based on duration statistics of the training datasets, we set $T'$ to 4 seconds, matching the average utterance length. To address class imbalance, we apply CE loss weights of 9 for bonafide and 1 for partial deepfakes. The model is trained using Adam \cite{kingma2014adam} with a learning rate of $5 \times 10^{-4}$, weight decay of $10^{-4}$, batch size of 2, for 10 epochs. The best model is selected based on validation loss. All experiments were performed on a single GeForce RTX 3090 GPU and the implementation code is publicly available at https://github.com/menglu-lml/inconsistency.

\section{Result and Discussion}
\subsection{Detection Results}

Table \ref{tab:RESULTS} shows that our proposed model, TDAM-AvgPool achieves the best results on both development and evaluation sets of the PS dataset. It outperforms state-of-the-art (SOTA) systems based on handcrafted features, learnable embeddings, and SSL-based representations by effectively capturing unnatural temporal fluctuations. Additionally, as shown in Table \ref{tab:RESULTS_HAD}, our model also achieves top performance on the HAD dataset, demonstrating high detection accuracy and strong generalization across languages and accent variations.

\begin{table}
\caption{Performance on the PS development and evaluation set in terms of EER for SOTA single systems and our proposed model.}
  \label{tab:RESULTS}
\renewcommand{\arraystretch}{1.1}
\begin{center}
\begin{tabular}{c|c|c|c}
\hline
\multirow{2}{*}{\textbf{Model}}&\multirow{2}{*}{\textbf{Front-end}}&\multicolumn{2}{|c}{\textbf{EER (\%)} $\downarrow$} \\
 \cline{3-4} 
&& Dev. Set& Eval. Set  \\
 \cline{1-4}
 Negroni et al. \cite{negroni24_asvspoof} &Spectrogram& 5.57 & 6.16 \\
 Khan et al. \cite{khan2024frame} &Spectrogram& - & 5.90 \\
 Zhu et al. \cite{zhu2023local} &LCNN encoder& - & 5.89 \\
 Cai et al. \cite{cai2024integrating} &wav2vec2-XLSR & 1.16 & 1.74 \\
 Liu et al. \cite{liu24m_interspeech}&wav2vec2-large & 0.35 & 0.73 \\
 Zhang et al. \cite{zhang2022partialspoof}&wav2vec2-large & 0.35 & 0.64 \\
 
 \textbf{TDAM-AvgPool (Ours)} & wav2vec2-XLSR&\textbf{0.19} & \textbf{0.59}\\
 \hline

\end{tabular}
\end{center}
\end{table}

\begin{table}
\caption{Performance on the HAD test set in terms of EER and AUC.}
  \label{tab:RESULTS_HAD}
\renewcommand{\arraystretch}{1}
\begin{center}
\begin{tabular}{c|c|c|c}
\hline
\multirow{2}{*}{\textbf{Model}}&\multirow{2}{*}{\textbf{Front-end}}&\multicolumn{2}{|c}{Test Set} \\
 \cline{3-4} 
& & \textbf{EER}(\%) $\downarrow$ & \textbf{AUC} $\uparrow$ \\
 \cline{1-4}
 Negroni et al. \cite{negroni24_asvspoof} &spectrogram& 7.36 & 95.24 \\
  Wu et al. \cite{wu2025weakly} &wav2vec2-XLSR& 4.56& 97.51 \\
    Wu et al. \cite{wu2024coarse} &wav2vec2-XLSR& 0.08& 99.96 \\
     Li et al. \cite{li2024wavlm} &wav2vec2& 0.07& - \\
 Cai et al. \cite{cai2024integrating} &WavLM  & 0.06 & 99.98 \\

 \textbf{TDAM-AvgPool (Ours)} & wav2vec2-XLSR&\textbf{0.03} & \textbf{99.99}\\
 \hline

\end{tabular}
\end{center}
\end{table}

\subsection{Ablation Studies}
% We perform ablation studies on the PartialSpoof evaluation set from three aspects to evaluate our proposed model.

\textbf{Study on design choices within TDAM} To assess the effectiveness of our model design, we progressively remove key components of TDAM-AvgPool: (1) removing the entire TDAM module, relying solely on SSL features (EER increases from 0.59\% to 0.87\%); (2) removing the dual-level structure of temporal difference representations (EER rises to 0.74\%); and (3) removing the second-level difference representation (EER at 0.80\%). As shown in the first three rows of Table~\ref{tab:ABLATION}, performance degrades in all cases. The largest drop occurs without TDAM, indicating that SSL features alone are insufficient for robust detection. Furthermore, replacing the raw difference representation with the absolute difference to eliminate directional variation leads to a 28.8\% drop in performance, highlighting the importance of preserving directional information. This demonstrates that frame-level directional changes are critical for modelling temporal dynamics in high-level acoustic features, while the dual-level design strengthens the representation by integrating multi-scale inconsistencies.

% \textbf{Study on comparing TDAM with other attention mechanisms} We replace TDAM with two widely used attention mechanisms: Squeeze-and-Excitation (SE) \cite{b25} and Efficient Channel Attention (ECA) \cite{b26}. As shown in Table \ref{tab:ABLATION}, both SE and ECA perform worse than TDAM. While SE and ECA effectively model global and local channel dependencies, they do not sufficiently capture unnatural temporal dynamics crucial for partial deepfake speech detection. This confirms that TDAM is explicitly designed to highlight unnatural variations over time, leading to superior detection performance over generic attention-based alternatives.

\begin{table}[t]
\caption{Ablation study results on the PS evaluation set with different design settings of TDAM.}
  \label{tab:ABLATION}
\renewcommand{\arraystretch}{1.05}
\begin{center}
\begin{tabular}{c|c}
\hline
\textbf{Method}  & \textbf{EER in Eval. set}  (\%) $\downarrow$ \\ 
\hline
 w/o TDAM &0.87\\
 w/o $\boldsymbol{X_1}$ and $\boldsymbol{X_2}$&0.74\\
  w/o $\boldsymbol{X_2}$ &0.80\\
w/o directional information& 0.76\\
% \hline
%  replace TDAM with SE &0.78\\
%  replace TDAM with ECA  &0.82\\

\hline
 TDAM-Trim-Pad  &1.03\\
 TDAM-MaxPool  &1.00\\
\hline
 \textbf{TDAM-AvgPool (Ours)} & \textbf{0.59}\\
 
 \hline

\end{tabular}
\label{tab1}
\end{center}
\end{table}

\begin{table}
\caption{Cross-dataset detection results in EER (\%), trained and tested on the ASVspoof 2019 LA and PS dataset}
  \label{tab:crossdataset}
\renewcommand{\arraystretch}{1.02}
\begin{center}

\begin{tabular}{l|cc|cc} 
\hline
& \multicolumn{2}{c|}{Training on LA}                                    & \multicolumn{2}{c}{Training on PS}                                    \\ \cline{2-5} & \multicolumn{1}{c|}{LA eval.}  & \multicolumn{1}{l|}{PS eval.} & \multicolumn{1}{l|}{LA eval.}  & \multicolumn{1}{l}{PS eval.} \\ \hline
\multicolumn{1}{c|}{Zhu et al. \cite{zhu2023local}}                 & \multicolumn{1}{c|}{3.09} & 13.79                             & \multicolumn{1}{c|}{5.24}          & 5.89                             \\ 
\multicolumn{1}{c|}{Zhang et al. \cite{zhang2022partialspoof}}                 & \multicolumn{1}{c|}{\textbf{0.83}} & 14.19                             & \multicolumn{1}{c|}{0.77}          & 0.64                             \\  \multicolumn{1}{c|}{\textbf{TDAM-AvgPool (Ours)}} & \multicolumn{1}{c|}{1.22}          & \textbf{11.42}                    & \multicolumn{1}{c|}{\textbf{0.71}} & \textbf{0.59} \\
\hline

\end{tabular}
\label{tab1}
\end{center}
\end{table}

\textbf{Study on handling varied-length inputs}  We compare our proposed approach for handling variable-length speech to two alternatives: TDAM-Trim-Pad and TDAM-MaxPool. TDAM-Trim-Pad standardizes utterances to 4 seconds via truncation or zero-padding. However, truncation may inadvertently remove critical deepfake regions, while simple concatenation distorts the original temporal flow, leading to a 42.7\% performance drop. TDAM-MaxPool replaces average pooling with max pooling, which overly emphasizes local peaks while failing to preserve the overall temporal structure, resulting in an EER of 1.00\%. These findings show that average pooling offers a better balance by aggregating directional variations across time, effectively preserving both strong and weak temporal cues and enabling robust detection across varied-length inputs.

% \begin{table}[t]
% \caption{Performance in EER (\%) for cross-dataset study. LA and PS stand for ASVspoof2019-
% LA and PartialSpoof datasets respectively.}
%   \label{tab:crossdataset}
% \renewcommand{\arraystretch}{1.1}
% \begin{center}
% \begin{tabular}{c|c|c}
% \hline
%  & LA evaluation set & PS evaluation set\\ 
% \hline
% Training on LA &1.22 & 11.42\\
% Training on PS & 0.71 & 0.59\\
% \hline

% \end{tabular}
% \label{tab1}
% \end{center}
% \end{table}

\subsection{Cross-dataset Study}
To evaluate generalization, we conduct a cross-dataset study using LA and PS datasets. The LA dataset consists of fully deepfake utterances, while PS includes partially deepfakes. As shown in Table~\ref{tab:crossdataset}, our model outperforms state-of-the-art models. The results indicate the strong generalizability of our model across diverse deepfake scenarios. Notably, its strong performance on fully deepfake samples suggests our model identifies artifacts within manipulated regions rather than relying on boundary transitions.
%However, we observe a performance drop when training on LA and testing on PS. This is expected, as the LA dataset lacks bonafide-deepfake segment comparisons and transitions, which limits the model’s ability to learn the distinct inconsistency patterns present in partial deepfakes.

\subsection{Visualization of Detected Temporal Irregularity Regions}
\begin{figure}
    \centering
    
    % First diagram
    \begin{minipage}{0.48\textwidth}
        \centering
        \includegraphics[width=\textwidth, height=0.65cm]{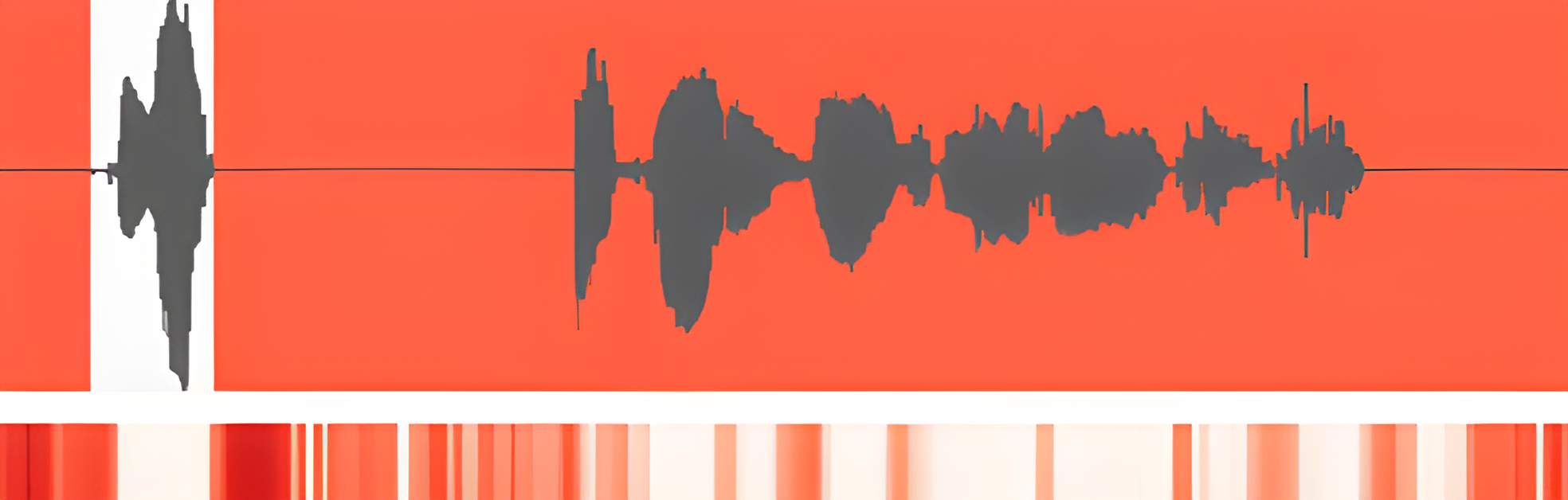}
        \vspace{1mm} % Adjust space if needed
        \footnotesize (a) CON\_E\_0012345.wav in PS evaluation set
    \end{minipage}
    
    % Second diagram
    \begin{minipage}{0.48\textwidth}
        \centering
        \includegraphics[width=\textwidth, height=0.65cm]{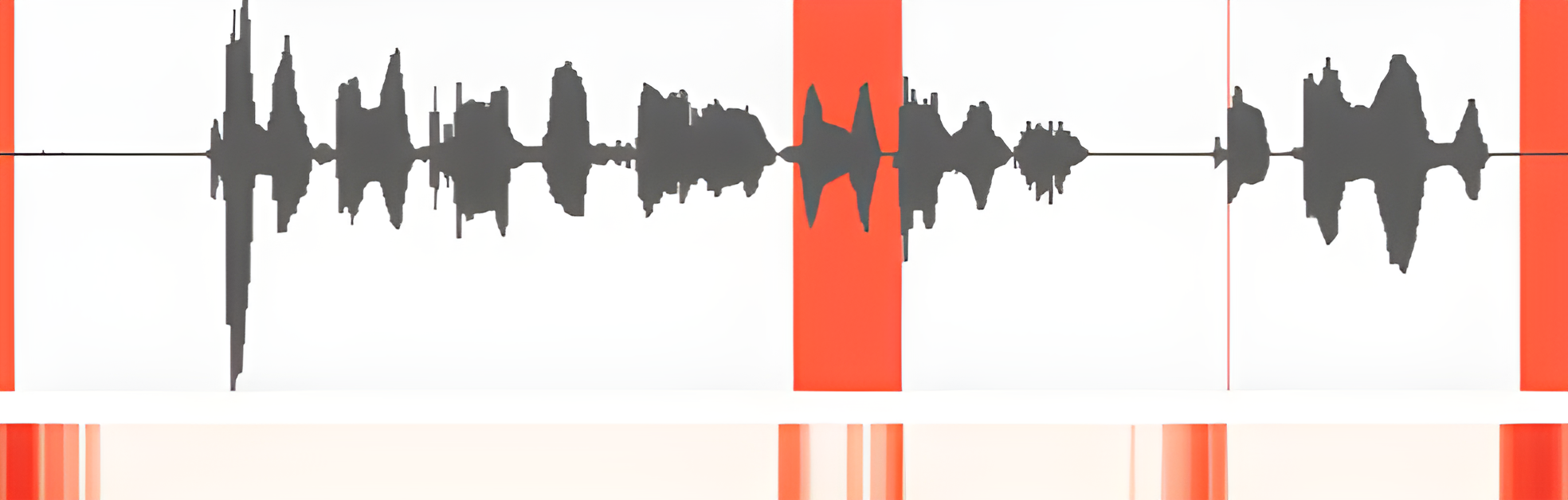}
        \vspace{1mm} % Adjust space if needed
        \footnotesize (b) CON\_E\_0023457.wav in PS evaluation set
    \end{minipage}

        \begin{minipage}{0.48\textwidth}
        \centering
        \includegraphics[width=\textwidth, height=0.65cm]{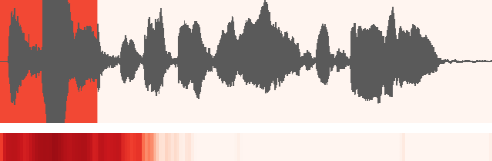}
        \vspace{1mm} % Adjust space if needed
        \footnotesize (c) HAD\_test\_00000074.wav in HAD test set
    \end{minipage}
    
    \caption{Visualization of ground truth and frame-level weighting maps for partial deepfake samples in the PS and HAD datasets. Red regions in the waveform indicate injected deepfake segments in the ground truth, while the heatmaps below highlight detected temporal irregularities. Darker shades represent higher confidence in identifying unnatural temporal variations.}
    \label{fig:heatmap}
\end{figure}

Fig. \ref{fig:heatmap} visualizes frame-level temporal irregularities detected by our model compared to ground truth labels. Red-highlighted regions in the waveform indicate deepfake segments, while white regions are bonafide. These deepfake segments are strategically injected at varying positions and durations. For each speech sample, the heatmap shows frame-level embeddings weighted by confidence scores, where darker shades correspond to stronger confidence in detecting unnatural temporal variations. This visualization demonstrates that despite training on utterance-level labels, TDAM-AvgPool accurately localizes artifacts not only at boundaries but also within deepfake regions. This suggests that our method captures underlying speech artifacts beyond simple boundary cues, making it more robust against advanced smoothing techniques that attempt to conceal transitions. Consequently, its ability to learn holistic temporal artifact patterns explains our model’s strong performance in detecting fully deepfake speech as well.

\section{conclusion}
This paper presents a novel approach to detecting partially deepfake speech by analyzing frame-level temporal differences across entire utterances. Motivated by the observation that deepfake speech exhibits unnatural temporal and directional variability compared to bonafide speech, we introduce a dual-level difference representation that effectively captures these irregularities. By leveraging temporal dynamics as indicators of synthetic content, our method eliminates reliance on explicit bonafide-deepfake transition boundaries and generalizes well to detect fully deepfake utterances. Trained exclusively with utterance-level labels, our TDAM-AvgPool model achieves state-of-the-art performance on both the PartialSpoof and Half-Truth datasets, demonstrating strong cross-lingual robustness.


\end{document}